\documentclass[12pt]{article}
\usepackage{graphicx}
\usepackage[utf8]{inputenc}
\usepackage{amsmath}
\usepackage{amsfonts}
\usepackage{amssymb}
\usepackage{graphicx}
\usepackage{color}

\begin{document}

\begin{center}
\Large {\bf  Electrodynamics under Action of Null Cosmic Strings}
\end{center}

\bigskip
\bigskip

\begin{center}
D.V. Fursaev and I.G. Pirozhenko
\end{center}

\bigskip
\bigskip

\begin{center}
{\it Dubna State University \\
     Universitetskaya st. 19\\
     141 980, Dubna, Moscow Region, Russia\\

  and\\

   Bogoliubov Laboratory of Theoretical Physics\\
  Joint Institute for Nuclear Research\\
  Dubna, Russia\\}
 \medskip
\end{center}

\bigskip
\bigskip

\begin{abstract}
A method to study  electromagnetic (EM) effects generated by a straight null cosmic string moving in classical EM fields is suggested. The  string is shown to induce an additional EM field which can be described as a solution to homogeneous Maxwell
equations with initial data set on a null surface, the string event horizon, where the string world-sheet belongs to. The initial data ensure the required holo\-nomy of the string space-time caused by the gravity of the string. This characteristic initial value problem is used to 
study the interaction of plane waves with null strings and perturbations by the strings of the Coulomb fields of electric charges.  It is shown that parts of an incoming EM wave crossing the string horizon from different sides of the string are refracted with respect to each other and leave behind the string a wedge-like region of interference. 
A null string moving near an electric charge results in two effects: it creates a self-force of the charge and induces a pulse of EM radiation traveling away from the charge in the direction close to trajectory of the string.
\end{abstract}

\newpage

\section{Introduction}\label{intr}

Cosmic strings \cite{Kibble:1976sj}, \cite{Vilenkin:2000jqa} are hypothetical astrophysical objects which might have been produced in the early Universe. Cosmic strings yield a variety of physical effects, such as lensing effects, the Kaiser-Stebbins effect \cite{Stebbins:1987va},  
which results in imprints of string's motion on cosmic microwave background \cite{Sazhina:2008xs}. The cusps of tensile strings emit strong beams of high-frequency gravitational waves \cite{Damour:2000wa} which may contribute to the stochastic gravitational background.
These effects are potentially observable \cite{Brandenberger:2013tr},\cite{LIGOScientific:2021nrg} and one expects that experimental evidences of cosmic strings would be an important step toward understanding physics at very high energies. 

Cosmic strings which appear as a result of the Kibble mechanism  \cite{Kibble:1976sj} are also called tensile strings since they have
a non-vanishing tension and nonzero rest mass per unit length.
A relatively less studied class of cosmic strings are null cosmic strings which are one-dimensional objects whose points move  along  trajectories of light rays, orthogonally to strings \cite{Schild:1976vq}.  The origin of null cosmic strings may be related to physics of
fundamental strings at the Planckian energies \cite{GM1}-\cite{Xu:2020nlh}. Equivalent names 
of null strings are massless strings \cite{vandeMeent:2012gb} or tensionless strings, to distinguish them from the tensile strings.
Like tensile strings 
null cosmic strings create holonomies of spacetime. The holonomies are null rotations belonging to the parabolic subgroup of the Lorentz group \cite{vandeMeent:2012gb}, \cite{Fursaev:2017aap}. The group parameter of the holonomies is determined by the string energy  per unit  length. 

Possible astrophysical and cosmological  effects  of null cosmic strings \cite{Fursaev:2017aap}, \cite{Fursaev:2018spa}, \cite{Davydov:2022qil}, such as deviations of light rays and trajectories of particles in the gravitational field of strings as well as scattering 
of strings by massive sources, are similar to that of the tensile strings. 
A distinctive feature of null strings is  their optical properties. The strings behave as one-dimensional null geodesic congruences  characterized by a complex scalar which is determined by 
an analogue of the Sachs' optical equation \cite {Fursaev:2021xlm}. The analysis shows that world-sheets of null strings develop caustics 
which accumulate large amounts of energy \cite{Davydov:2022qil}.

The main purpose of the present paper is to describe new electromagnetic  (EM) effects generated by a straight null string in locally Minkowsky space-time.  On the technical side our aim is to define  classical electrodynamics on 
space-times with  null holonomy. It should be noted that field theories in the background geometry of a straight tensile string
have been studied earlier in numerous publications in a reference frame  where the string is at rest 
and the corresponding space-time has conical singularities, see e.g. \cite{Dowker:1977zj}-\cite{Davies:1987th} 
among the pioneering papers. Field theories in the presence of null strings is relatively unknown research area.

A approach to physical effects caused by null strings 
has been suggested in \cite{Fursaev:2017aap}. In the string space-time the null holonomy transformations
have fixed points on the string world-sheet which belongs to a null hypersurface ${\cal H}$, the string event horizon.
The idea of \cite{Fursaev:2017aap} is to set, for matter crossing the string horizon, ``initial'' data on ${\cal H}$ to ensure the
required holonomy transformations. The approach has been used in \cite{Fursaev:2017aap}, \cite{Fursaev:2018spa} to describe the Kaiser-Stebbins effect caused by null cosmic strings.

In the present paper, we  extend the method of \cite{Fursaev:2017aap} to study
observable effects  of classical  EM fields generated by null cosmic strings.

The paper is organized as follows.  In Sec.  \ref{holo} we describe the holonomy method of \cite{Fursaev:2017aap}  
with the focus on free field theories. Finding solutions to wave equations in 
a free field theory on a space-time of a null string, in the domain above
${\cal H}$, is equivalent  to solving an initial value problem with initial data on ${\cal H}$ determined by incoming data. The initial data are null rotated so that to ensure the required holonomy.  In the theory of hyperbolic second order  partial differential equations (PDE) such an initial value (Cauchy) problem is called  the characteristic initial value problem \cite{Morse:1953} since the standard pair of initial data are not independent on $\cal H$. We discuss in details how the
initial data should be chosen in the  case of scalar field theory and in Maxwell's theory, where gauge constraints should be taken into account. We also demonstrate that observers crossing $\cal H$ do not see discontinuities in the stress-energy tensor of the fields. 

In next Sections we apply this method to study different EM effects induced by null strings.
Scattering of 
monochromatic plane electromagnetic waves on null strings is considered in Sec \ref{refraction}.  The null cosmic string cuts the wave front into two waves and changes on $\cal H$ their directions  relative to each other. That is, the string horizon acts as a refractive media. As a result, the string leaves behind a wedge-like region, an interference wedge, where the refracted waves interfere.  
An observer inside the interference wedge sees
two waves, as if they come from different sources.
This property is the manifestation of the lensing effect. The interference 
is also analogous to creation by cosmic strings of overdensities of matter.
In Sec. \ref{ED} we define EM fields created by charges in the presence of the straight null string. Solution of the Maxwell equations
above the string horizon ${\cal H}$ is studied in detail for a single point charge. 
We show that in addition to standard Coulomb field of the charge the string generates rapidly changing EM field which acts as a self-force on the charge. 
 At large times the additional field looks as a pulse of EM radiation traveling away from the charge.
We calculate numerically the energy density and the energy flow of the pulse  and show that its duration and peak is determined 
by the impact parameter between the string and the charge.
Other  applications of our results are discussed in Sec. \ref{appl}. The suggested method can be used to describe classical field effects on gravitational shock-wave backgrounds. Section \ref{sum} is a summary.
Our analysis is based on exact solution for the Cauchy problem for a scalar plane wave. The derivation of this solution and the properties 
of the Green function with a delta-function source on $\cal H$ can be found in Sec. \ref{App1}.
Some details regarding the homogeneous solution for the Coulomb potential  are given in Sec. \ref{App2}.

\section{Field Dynamics and Null Holonomies}\label{holo}
\setcounter{equation}0

\subsection{Coordinate conditions on the string horizon}\label{charts}

We consider a straight cosmic string which is stretched along $z$-axis and moves along $x$-axis in $R^{1,3}$. 
It is convenient to use the light-cone coordinates  $v=t+x$, $u=t-x$, where the metric is
\begin{equation}\label{i.1a}
ds^2=-dv du +dy^2+dz^2~~.
\end{equation} 
The string world-sheet can be defined by equations $u=y=0$. 
The parabolic subgroup of the Lorentz transformations (null rotations), $(x')^\mu=M^\mu_{~\nu}\left(\lambda\right)x^\nu$, acts on $u,v,y,z$ coordinates in $R^{1,3}$ as follows:
\begin{equation}\label{1.1}
u'=u~~,~~
v'=v+2\lambda y+\lambda^2u~~,~~
y'=y+\lambda u~~,~~z'=z~~,
\end{equation}  
where $\lambda$ is some real parameter.  Transformation of a vector is
\begin{equation}\label{1.1a}
V_u'=V_u-\lambda V_y+\lambda^2 V_v~~,~~
V_v'=V_v~~,~~
V_y'=V_y-2\lambda V_v~~,~~V_z'=V_z~~,
\end{equation}
or  $V'_\mu=M_{\mu}^{~\nu}(\lambda )V_\nu$, where $M_{\mu}^{~~\nu}=\eta_{\mu\mu'}\eta^{\nu\nu'}M^{\mu'}_{~~\nu'}$.

For a null string with the world-sheet $u=y=0$ a parallel transport of a vector $V$ along a closed contour around the string 
results in a null rotation, $V'=M(\omega)V$ with $\omega$ defined as, see \cite{vandeMeent:2012gb},
\begin{equation}\label{i.2}
\omega \equiv 8\pi GE~~.
\end{equation} 
The world-sheet is a fixed point set of (\ref{1.1}). The hypersurface $u=0$ is the event horizon of the string. We denote  it
by ${\cal H}$.

The holonomy method suggested in \cite{Fursaev:2017aap} is to set initial data on the string horizon.
To determine these data the string space-time is decomposed onto two parts: $u<0$, and $u>0$.  We call trajectories of particles and light rays at $u<0$ and $u>0$ 
ingoing and outgoing trajectories, respectively. 
To describe outgoing trajectories, one introduces two types 
of coordinate charts: $R$- and $L$-charts, with cuts on the horizon either on the left
($u=0, y<0$) or on the right ($u=0, y>0$)  to the string, respectively. The initial data on the string horizon are related to the ingoing data via null rotations (\ref{1.1}) taken at $u=0$. For  brevity the right  ($u=0, y>0$) and the left ($u=0, y<0$) parts of $\cal H$ will be denoted as ${\cal H}_+$
and ${\cal H}_-$, respectively.

For the $R$-charts the cut is along ${\cal H}_-$. The coordinate transformations
on the string horizon and  the initial data for the outgoing trajectories  are
\begin{equation}\label{1.3}
x_+^\mu=\bar{x}^\mu\mid_{{\cal H}_+}~~,
~~u_+^\mu=\bar{u}^\mu\mid_{{\cal H}_+}~~,
\end{equation}  
\begin{equation}\label{1.2}
x_-^\mu=M^\mu_{~\nu}\left(\omega\right)\bar{x}^\nu\mid_{{\cal H}_-}~~,
~~u_-^\mu=M^\mu_{~\nu}\left(\omega\right)\bar{u}^\nu\mid_{{\cal H}_-}~~.
\end{equation}  
where $\bar{x}^\mu$, $\bar{u}^\mu$ are the coordinates and velocities of the corresponding ingoing trajectory when it reaches the horizon.   
It follows from  (\ref{1.1}) that the coordinate transformations (\ref{1.3}) at $y<0$ are reduced to a shift of a single coordinate:
\begin{equation}\label{1.3a}
v_-=\bar{v}+2\omega y~~,~~y<0~~.
\end{equation} 
Thus, on the $R$-charts the 'right' trajectories ($y>0$) behave smoothly across the horizon, while the `left' trajectories  ($y<0$) are shifted along the $v$ coordinate and change their direction under the null rotation.

The descriptions based on $R$- or $L$-charts are equivalent.
The choice of a chart is a matter of convenience depending on the observer's trajectory.
$L$-charts are dual to $R$-charts. They are smooth everywhere except the 
right cut on the horizon, ${\cal H}_+$.  The shift of the coordinate now is
\begin{equation}\label{1.3d}
v_+=\bar{v}-2\omega y~~,~~y>0~~.
\end{equation} 
'Right' outgoing trajectories experience null rotations like in (\ref{1.3}), (\ref{1.2}) where 
$\omega$ should be replaced with $-\omega$.

The reason why descriptions in terms of $R$- and $L$-charts are equivalent is that only relative transformations of 
'left' and 'right' outgoing trajectories have physical and geometrical meanings.

\subsection{Characteristic Cauchy problem for scalar fields}\label{Cauchy}

The above holonomy method can be extended to describe classical field theories, or fibre bundles over the null string geometry.
In this Section we consider non-interacting scalar fields $\phi$ with equation
\begin{equation}\label{1.5}
(\Box -m^2)\phi(x)=j(x)~~,
\end{equation} 
\begin{equation}\label{i.1b}
\Box=\partial_\mu\partial^\mu=-4\partial_u \partial_v+\partial^2_y+\partial^2_z~~,
\end{equation} 
where $j(x)$ is an external source. 
It is a second order hyperbolic PDE which allows a well-posed Cauchy problem on initial  space-like  hypersurfaces.
The initial data include fields and their first time derivatives. 

Since we need  solutions of (\ref{1.5}) above  $\cal H$, $u>0$, it is natural to consider the initial value problem with $\cal H$ as the initial
hypersurface. The point is that $\cal H$ is null, and it is a characteristic surface \cite{Morse:1953}  of (\ref{1.5}), where standard Cauchy data, fields and their first derivatives, are not  independent.  A solution of (\ref{1.5}) can be fixed just by the value of the field  
$\hat{\phi}({\bf x})=\phi(x)\mid_{\cal H}$,  where ${\bf{x}}\equiv(v,y,z)$.  An analogue of a time derivative, $\chi({\bf{x}})=\partial_u \phi(x)$ at $u=0$, can be expressed 
from (\ref{1.5}) as
\begin{equation}\label{td}
\chi({\bf{x}})=\frac 14\int_{-\infty}^v dv' \left[(\partial_y^2+\partial_z^2-m^2)\hat{\phi}({\bf x}')-j({\bf x}')\right]
+f(y,z)~~,
\end{equation} 
where ${\bf{x}}'\equiv(v',y,z)$. Asymptotic properties of $\chi({\bf{x}})$ at future or past null infinities
depend on an arbitrary function $f(y,z)$ in (\ref{td}). This function
can be fixed or eliminated by requiring appropriate behavior of $\hat{\phi}({\bf x})$ and $\chi({\bf{x}})$
at null infinities. Under these conditions $\chi$ is fixed by $\hat{\phi}$ and $j$ on $\cal H$.

To take into account the holonomy of the string space-time the initial data on ${\cal H}_+$ and ${\cal H}_-$ should be considered
separately. We denote them as
\begin{equation}\label{1.3ds}
\phi(x)\mid_{{\cal H}_\pm}=\hat{\phi}_\pm({\bf x})~~.
\end{equation} 
In the $R$-chart, the continuity of solutions across $\cal H$ is ensured by the following transition conditions:
\begin{equation}\label{1.3cs}
\hat{\phi}_+({\bf x})=\bar{\phi}({\bf x})\mid_{{\cal H}_+}~~,~~
\hat{\phi}_-({\bf x})=\bar{\phi}(\bar{\bf x})\mid_{{\cal H}_-}~~,
\end{equation}  
$$
\bar{\bf x}={\bf x}-2\omega y{\bf{q}}
$$
where $\bar{\phi}$ is the value on $\cal H$ of the ingoing field at $u<0$, and $q^i=\delta^i_v$.  We also require continuity of the current
in the r.h.s. of (\ref{1.5}),
\begin{equation}\label{1.3cur}
j({\bf x})\mid_{{\cal H}_+}=\bar{j}({\bf x})~~,~~
j({\bf x})\mid_{{\cal H}_-}=\bar{j}(\bar{\bf x})~~.
\end{equation}
Conditions (\ref{1.3cs}) are analogous to conditions (\ref{1.3}), (\ref{1.2})  for trajectories of particles and light rays.
According to (\ref{1.3cs}) a left observer with `ingoing' coordinates $\bar{\bf{x}}$ on ${\cal H}_-$ will be shifted to a coordinate 
$\bar{\bf{x}}+2\omega y{\bf{q}}$. The transition condition (\ref{1.3cs}) means that the left observer measures in the new coordinates the same value of 
the field and the current. 

Let us show that vector field $V_\mu=\partial_\mu \phi$ has the required holonomy when going around the string world-sheet. To this aim 
it is enough to demonstrate that 'outgoing' and 'ingoing' data on $\cal H$, in the $R$-chart, are related as
\begin{equation}\label{2.5s}
V_{\mu}({\bf x})=\bar{V}_\mu(x)\mid_{{\cal H}_+}~~,~~
V_{\mu}({\bf x})=M_{\mu}^{~\nu}(\omega)\bar{V}_\nu(\bar{x})\mid_{{\cal H}_-}~~.
\end{equation}
Here $M_{\mu}^{~\nu}(\omega)$ are defined in (\ref{1.1a}).  It is easy to see that (\ref{1.1a}) is fulfilled on ${\cal H}_-$ for $v$ and  $z$ components. 
For the $y$ component one finds
\begin{equation}\label{2у}
\partial_y\phi (x)\mid_{{\cal H}_-}=(\partial_{\bar{y}}-2\omega  \partial_{\bar{v}})\bar{\phi}(\bar{\bf x})\mid_{{\cal H}_-}~~,
\end{equation} 
in agreement with (\ref{1.1a}).  
The result for $\chi({\bf{x}})=\partial_u \phi(x)$ at ${\cal H}_-$ follows from (\ref{1.5}) and (\ref{2у}) which imply
$$
\partial_v\chi({\bf{x}})=\frac 14\left[(\partial_y^2+\partial_z^2-m^2)\phi({\bf x})-j({\bf x})\right]=
$$
\begin{equation}\label{2u}
\frac 14\left[((\partial_{\bar{y}}-2\omega  \partial_{\bar{v}})^2+\partial_{z}^2-m^2)\bar{\phi}(\bar{\bf x})-\bar{j}(\bar{\bf x})\right]=\partial_{\bar{v}}\left(\bar{\chi}(\bar{\bf{x}})-\omega \partial_{\bar{y}}\bar{\phi}(\bar{\bf x})+\omega^2 \partial_{\bar{v}}\bar{\phi}(\bar{\bf x})\right)~~,
\end{equation} 
where $\bar{\chi}(\bar{\bf{x}})=\partial_u \bar{\phi}(\bar{x})$ at ${\cal H}_-$. That is, under corresponding conditions at 
null infinities,
\begin{equation}\label{3u}
\chi({\bf{x}})=\bar{\chi}(\bar{\bf{x}})-\omega \partial_{\bar{y}}\bar{\phi}(\bar{\bf x})+\omega^2 \partial_{\bar{v}}\bar{\phi}(\bar{\bf x})~~.
\end{equation}
This coincides with null rotation of the $u$ component in  (\ref{1.1a}).  It is clear that relation (\ref{3u}) is a consequence of relativistic 
invariance of equation of motion (\ref{1.5}).

The stress-energy tensor of the fields $T_{\mu\nu}$ is constructed of $\phi$ and $\partial_\mu \phi$. Since
\begin{equation}\label{2.5t}
T_{\mu\nu}(x)=M_{\mu}^{~\alpha}(\omega)M_{\nu}^{~\beta}(\omega)\bar{T}_{\alpha\beta}(\bar{x})\mid_{{\cal H}_-}~~,
\end{equation} 
it has the required holonomy which belongs to tensor representation of the Lorentz group. Transition conditions (\ref{2.5t}) guarantee that left observers with 4-velocities $u_o$  do not see discontinuities  in quantities like 
$u_o^\mu \partial_\mu \phi$ or $u_o^\mu u_o^\nu T_{\mu\nu}$  when crossing the string horizon.  

\bigskip

We can now formulate the characteristic initial value problem:

\begin{itemize}
\item[--] the solutions $\phi(x)$ of hyperbolic type PDE (\ref{1.5}) are looked for in the domain $u>0$;

\item[--] the initial data are set on null hypersurface $\cal H$ ($u=0$) and consist of a single variable, the value of the field $\hat{\phi}({\bf x})$, appropriate asymptotic conditions at null infinities are assumed;

\item[--]  the initial data $\hat{\phi}({\bf x})$ are determined by `incoming' solution $\bar{\phi}(\bar{x})$ in the domain $u<0$ with the help of transition conditions (\ref{1.3cs}) which are synchronized with coordinate conditions (\ref{1.3}), (\ref{1.2}).
\end{itemize}

A solution to (\ref{1.5}),(\ref{1.3cs}) can be written as 
\begin{equation}\label{1.8}
\phi(x)=\phi_I(x)+\phi_H(x)~~.
\end{equation}
Here $\phi_I(x)$ is a particular solution to inhomogeneous equation (\ref{1.5}) in $R^{1,3}$ taken at $u>0$. 
We denote by $\hat{\phi}_{I,\pm}$ the corresponding data of  $\phi_I(x)$ on ${\cal H}_\pm$. 
The field $\phi_H(x)$ is a solution to a homogeneous problem
\begin{equation}\label{1.9}
(\Box -m^2)\phi_H(x)=0~~,~~\phi_H(x)\mid_{{\cal H}_\pm}=\hat{\phi}_{H,\pm}({\bf x})~~,~~\hat{\phi}_{H,\pm}({\bf x})=\hat{\phi}_{\pm}({\bf x})-
\hat{\phi}_{I,\pm}({\bf x})~~.
\end{equation} 
The Cauchy data in (\ref{1.9}) are chosen so that to ensure the required data (\ref{1.3cs}) for $\phi(x)$. 
The solution to (\ref{1.9})  can be written as 
\begin{equation}\label{1.6}
\phi_H(x)=\int_{y'>0}d{\bf x'}D(u,{\bf x-x'})\hat{\phi}_{H,+}({\bf x}')+\int_{y'<0}d{\bf x'}D(u,{\bf x-x'})\hat{\phi}_{H,-}({\bf x}')~~,
\end{equation} 
where the $D$-function is the solution to the following problem:
\begin{equation}\label{1.7}
(\Box -m^2) D(x)=0~~,~~D(u,{\bf x})\mid_{u=0}=\delta^{(3)}({\bf x})~~.
\end{equation}

It is convenient to assume that $\phi_I$ is the solution when the string is absent. Then physical effects generated by the null string are related 
to $\phi_H$. It is this homogeneous part  we study  in next Sections  in different physical situations.

\subsection{Characteristic Cauchy problem for a Maxwell field}\label{Cauchy-EM}

Our primary interest is electromagnetic fields on space-time of a null cosmic string.
The standard initial value problem for the Maxwell equations in Minkowsky space-time,
\begin{equation}\label{em.1}
\partial_\mu F^{\mu\nu}=j^\nu~~,
\end{equation}
$F_{\mu\nu}=\partial_\mu A_\nu - \partial_\nu A_\mu$, $\partial j=0$,
with initial {\it space-like} hypersurface $t=0$ is determined by the following initial data:
\begin{equation}\label{em.2}
A_i\mid_{t=0}=a_i~~,~~F_{0i}\mid_{t=0}=\pi_i~~,~~i=x,y,z~~.
\end{equation}
The pairs $a_i,\pi_i$ are canonical coordinates and momenta. The gauge symmetry
imposes the constraint
\begin{equation}\label{em.3}
\partial_i \pi_i=-j_0\mid_{t=0}~~,
\end{equation}
which leaves 2 independent momenta. Also the gauge transformations, $\delta A_\mu =\partial_\mu \lambda$, allow one to exclude one of coordinates $a_i$.
Therefore there are only 4 independent initial data.

For the subsequent analysis, it is convenient to use the Lorentz gauge condition $\partial A=0$  since it is invariant under holonomy transformations
on $\cal H$ and can be imposed globally on cosmic string space-time. The Maxwell equations are reduced to
\begin{equation}\label{2.1}
\Box A_\mu=j_\mu~~,~~\partial A=0~~.
\end{equation}
The initial data for (\ref{2.1}) are
\begin{equation}\label{em.6}
A_\mu \mid_{t=0}=a_\mu~~,~~\dot{A}_\mu\mid_{t=0}=p_\mu~~.
\end{equation}
The data $a_0$ and $p_0$ are not independent:  $p_0$ is fixed by the gauge condition, while $a_0$  is 
determined by constraint (\ref{em.3}).
The gauge freedom of (\ref{2.1}), $\delta A_\mu =\partial_\mu \lambda$, $\Box \lambda=0$, that implies transformations of the  initial data, 
\begin{eqnarray*}
\delta a_i =\partial_i \lambda, \quad  \delta a_0=\dot{\lambda}, \quad \delta p_i =\partial_i \dot{\lambda}, \quad \delta p_0=\triangle \lambda,
\end{eqnarray*} leaves 4 independent 
initial data.

Consider now the {\it characteristic initial value problem} for Maxwell equations with initial hypersurface $\cal H$ ($u=0$). The canonical coordinates
and momenta can be determined by using the Hamilton-Jacobi method from the variation of the Maxwell action with $\cal H$ as
a boundary,
\begin{equation}\label{em.4}
A_b\mid_{\cal H}=a_b~~,~~F^{ub}\mid_{\cal H}=\pi_b~~,~~b=v,y,z~~.
\end{equation}
The data are subject to the constraint
\begin{equation}\label{em.5}
\partial_v \pi_v+\partial_x \pi_x+\partial_y \pi_y=-j^u\mid_{\cal H}~~.
\end{equation}
The momenta  $\pi_y=2F_{vy}$, $\pi_z=2F_{vz}$ are completely determined by the initial data $a_b$. If these momenta are known, $\pi_v=4F_{uv}$
is fixed by (\ref{em.5}). Since there is the gauge freedom $\delta a_b =\partial_a \lambda$ in definition of $\pi_y,\pi_z$,
the initial value problem on $\cal H$ requires 2 independent data, twice less than that in the initial problem on space-like hypersurface.

If we fix the Lorentz gauge the initial data for (\ref{2.1}) can be formulated as
\begin{equation}\label{em.7}
A_\mu \mid_{\cal H}=a_\mu~~,
\end{equation}
where $a_u$ should be determined by the gauge condition.  The remaining gauge freedom then leaves 2 independent data.
Constraint (\ref{em.5}) follows from  the gauge condition and equation for the $v$ component.

We go now to Maxwell equations on string space-time. Let $\bar{A}_\mu$ be a solution to (\ref{em.1}) with an electric 
current  $\bar{j}^\mu$ 
in the region $u<0$. We continue to work in the $R$-chart. By analogy with (\ref{1.3cur}), the following conditions 
ensure continuity of the current across $\cal H$:
\begin{equation}\label{em.8}
j^\mu({\bf x})\mid_{{\cal H}_+}=\bar{j}^\mu({\bf x})~~,~~
j^\mu({\bf x})\mid_{{\cal H}_-}=M^{\mu}_{~\nu}(\omega)\bar{j}^\nu(\bar{\bf x})~~,
\end{equation}
where $j^\mu$ is defined at $u>0$, and $\bar{\bf x}={\bf x}-2\omega y{\bf{q}}$.  The conservation law, $\partial j=0$, is invariant 
with respect to null rotations. Together with (\ref{em.8}) it implies that the electric charge $Q$ defined on null surfaces $u=C$,
\begin{equation}\label{3.6}
Q=\int_{u=C}d\Sigma_\mu j^\mu(u,{\bf x})=\int_{u=C} j^u(u,{\bf x})dv dy dz~~
\end{equation}
does not change when crossing $\cal H$.

The characteristic initial value problem for EM fields at $u>0$ includes field equations (\ref{2.1})
and initial data (\ref{em.7}) set on $\cal H$. The data are determined as
\begin{equation}\label{2.5}
a_{b}({\bf x})=\bar{a}_b({\bf x})\mid_{{\cal H}_+}~~,~~
a_{b}({\bf x})=M_{b}^{~c}(\omega)\bar{a}_c(\bar{\bf x})\mid_{{\cal H}_-}~~,
\end{equation}
\begin{equation}\label{em.9}  
\bar{a}_b ({\bf x})=\bar{A}_b (x)\mid_{{\cal H}}~~,~~b=v,y,z~~.
\end{equation}
Note that $M_{b}^{~u}=0$, see (\ref{1.1a}). The initial data for the $u$-component is determined by the gauge condition
$\partial A=0$. By taking into account  (\ref{2u}) and the fact that $a_v$ is invariant under null rotations  one finds that
\begin{equation}\label{em.10}
\partial_v a_u=\partial_{\bar{v}}(\bar{a}_u-\omega \bar{a}_y+\omega^2 \bar{a}_v)~~.
\end{equation}
This relation holds on ${\cal H}_-$ and is consistent with the transformation law $a_u({\bf x})=M_u^{~\mu} \bar{a}_\mu (\bar{\bf x})$. By using this one
can demonstrate the transition conditions for the Maxwell tensor on ${\cal H}_-$ 
\begin{equation}\label{em.11}
F_{\mu\nu}(x)=M_{\mu}^{~\alpha}(\omega)M_{\nu}^{~\beta}(\omega)\bar{F}_{\alpha\beta}(\bar{x})\mid_{{\cal H}_-}~~,
\end{equation}
in accord with the holonomy of the space-time.

As is explained in Sec. \ref{Cauchy} it is convenient to look for a solution to (\ref{2.1}), (\ref{2.5}) in the form:
\begin{equation}\label{3.7}
A_\mu(x)=A_{I,\mu}(x)+A_{H,\mu}(x)~~,
\end{equation}
where $A_{I,\mu}$ is a particular solution to (\ref{2.1}) and 
$A_{H,\mu}$ is a solution to a homogeneous characteristic initial value problem
\begin{equation}\label{3.8a}
\Box A_{H,\mu}=0~~,~~\partial A_H=0~~,~~A_{H,b} (x)\mid_{{\cal H}}=a_{H,b}({\bf x})~~.
\end{equation} 
If we assume that $A_{I,\mu}$ coincides with the solution in the absence of the string, that is $A_{I,\mu}=\bar{A}_{\mu}$ on ${\cal H}_+$,
the initial data in (\ref{3.8a}) become
\begin{equation}\label{3.8b} 
a_{H,b}({\bf x})\mid_{{\cal H}_+}=0~~,~~a_{H,b}({\bf x})\mid_{{\cal H}_-}=M_{b}^{~c}(\omega)\bar{a}_c(\bar{\bf x})-a_{I,b}({\bf x})
~~,
\end{equation} 
where  $a_{I,b}({\bf x})=A_{I,b} (x)\mid_{{\cal H}}=\bar{a}_b({\bf x})$. 

\section{Refraction of EM waves on the string horizon}\label{refraction}
\setcounter{equation}0

The first type of physically interesting effects is related to scattering of electromagnetic waves on null strings.
Consider monochromatic plane waves which have the standard form before scattering on the string (in the region $u<0$): 
\begin{equation}\label{2.3}
\bar{A}_\mu(\bar{x})=\Re~(\bar{E}_\mu~e^{i \bar{k}\cdot \bar{x}})~~,
\end{equation}
where $\bar{E}_\mu$ is some complex polarization vector, $\bar{k}^\mu\bar{E}_\mu=0$. As earlier, we denote the incoming data with the bar.
Other types of electromagnetic waves can be treated as a superposition of plane monochromatic waves. 

We are dealing with (\ref{2.1}) when $j=0$. This simplifies the choice of data on $\cal H$.

On the R-chart there is no transformation of the part of the wave crossing ${\cal H}_+$.
If (\ref{2.5}) are applied to (\ref{2.3}) on ${\cal H}_-$ one concludes 
that the wave leaves ${\cal H}_-$ with the transformed momentum
\begin{equation}\label{2.7}
k_-^\mu=M^\mu_{~\nu}\left(\omega\right)\bar{k}^\nu~~.
\end{equation}
The transformed momentum $k_-$ is introduced to satisfy the condition $\bar{k}\cdot \bar{x}\mid_{{\cal H}_-}=k_-\cdot x_-$,
where $x_-$ are defined by (\ref{1.2}).  The initial data (\ref{2.5}), (\ref{em.9}) and the gauge  $\partial A=0$
then imply the following conditions: 
\begin{equation}\label{2.23a}
E^+_{\mu}=\bar{E}_\mu~~,~~E^-_{\mu}=M_{\mu}^{~\nu}(\omega)\bar{E}_\nu~~
\end{equation}
for the polarization vectors of waves which leave ${\cal H}_+$ and ${\cal H}_-$, respectively.

For the right observers the wave from ${\cal H}_-$ changes its energy and looks refracted. 
If $E$ and $\vec{k}$ are, respectively, the energy and the momentum of the incoming wave,
the  refraction angle  $\varphi_{\mbox{\tiny{refr}}}$ and the energy of the refracted wave are
\begin{equation}\label{2.8}
\cos\varphi_{\mbox{\tiny{refr}}}={(\vec{k}_-\vec{k}) \over E_-E}={1 \over EE_-}\left[E^2+{\omega^2 \over 2}(Ek^x-(k^x)^2)+\omega Ek^y\right]
\end{equation}
\begin{equation}\label{2.9}
E_-=\left(1+{\omega^2 \over 2}\right)E-{\omega^2 \over 2}k^x+\omega k^y~~.
\end{equation}
The refraction is absent only for the waves traveling along the string axis $z$ when $k^x=k^y=0$.

To study physical effects caused by the refraction of waves on $\cal H$
we need to solve (\ref{2.1}) with the Cauchy data (\ref{2.5}).  Since the problem is homogeneous 
its solution is given by (\ref{1.6}). In case of the monochromatic plane waves
each component of $A_\mu$ can be treated as a scalar wave. If one ignores the effects related to polarizations, the
basic features of the scattering problem can be understood by studying a scalar field theory with equation 
\begin{equation}\label{2.10}
\Box \phi=0~~.
\end{equation}
Suppose that the scalar field $\phi$ behaves at $u<0$ as
\begin{equation}\label{2.11}
\bar{\phi}(\bar{x})=e^{i \bar{k}\cdot \bar{x}}~~.
\end{equation}
In the domain $u>0$  the scattered wave (\ref{2.11}) is a superposition,
$$
\phi(x)=\int_{y'>0}d{\bf x'}D(u,{\bf x-x'})e^{i k_+\cdot x'}\mid_{u'=0}+\int_{y'<0}d{\bf x'}D(u,{\bf x-x'})e^{i k_-\cdot x'}\mid_{u'=0}
$$
\begin{equation}\label{2.12}
\equiv \phi_+(x)+\phi_-(x)~~,
\end{equation}
where $k_+=\bar{k}$, $k_-=M(\omega)\bar{k}$. To represent solutions $\phi_\pm(x)$ we introduce the following dimensionless functions:
\begin{equation}\label{2.13}
f(k,x)=u k_y+2 k_v y~~,~~g(k,x)=\frac{f^2(k,x)}{4k_v u}~~,
\end{equation}
\begin{equation}\label{2.14}
f_\pm=f(k_\pm,x)~~,~~g_\pm=g(k_\pm,x)~~.
\end{equation}
After some algebra one gets, see Sec. \ref{App1},
\begin{equation}\label{2.15a}
\phi_\pm(x)=[\theta(\pm f_{\pm})+\varepsilon(\pm f_{\pm}) G(g_\pm)]\exp(i k_{\pm} x)~~,~~k_v>0
\end{equation}
\begin{equation}\label{2.15b}
\phi_\pm(x)=[\theta(\mp f_{\pm})+\varepsilon(\mp f_{\pm}) G^*(-g_\pm)]\exp(i k_{\pm} x)~~,~~k_v<0~~.
\end{equation}
where $\theta$ and $\varepsilon$ are the step and the sign functions, respectively.
The complex factor $G(g)$ is defined, for $\Re ~g>0$, as
\begin{equation}\label{2.16}
G(g)=-\frac{e^{i\pi/4}}{\pi}\int_{0}^{\infty}\frac{dt}{t^2+i}e^{-g(t^2+i)}=-\frac{1}{2}\mbox{Erfc}(\sqrt{i g})~~.
\end{equation}
By using  (\ref{2.13})	-(\ref{2.16}) one can check that $\Box \phi_\pm=0$ at $u>0$.

The $G$-factor has the following expansions at small and large $g$:
\begin{eqnarray}
G(g)&=&-\frac{1}{2}+\frac{e^{i\pi/4}}{\sqrt{\pi}} g^{1/2}\left(1-\frac{i}{3}g+\dots\right), \quad g\to 0,
\label{asympt_zero}\\
G(g)&=&\frac{e^{ i \pi /4}}{2\sqrt{\pi}}\,  e^{-i g} g^{-1/2} 
\left(e^{ i \pi /2}   -\frac{1}{2 g} +\dots           \right), \quad g\to \infty.
\label{asympt_inf}
\end{eqnarray}
As a consequence of (\ref{asympt_inf}), $G$ vanishes as $u\to 0$, and $\phi_\pm$ satisfy the required boundary conditions.
One can also check with the help of (\ref{asympt_zero}) that solutions (\ref{2.15a}), (\ref{2.15b}) are continuous across the surfaces 
$f_\pm=0$.

Due to the presence of the $G$-factor the scattered wave is not monochromatic  near the string world-sheet, $u\to 0$,  $g\to 0$, in a 'near-field zone'. In a 'far-field zone', $g\gg 1$, the wave has a simple form.
For instance, for $k_v>0$, it is
\begin{equation}\label{2.17a}
\phi(x)=\theta(f_{+})\exp(i k_{+} x)+\theta(-f_{-})\exp(i k_{-} x)+\phi_{\mbox{\tiny{tail}}}(x)~~,
\end{equation}	
\begin{equation}\label{2.17b}
\phi_{\mbox{\tiny{tail}}}(x)=\varepsilon(f_{+}) {1 \over \sqrt{4\pi g_+}}e^{i k_{+} x+i\varphi_+}+
\varepsilon(-f_{-}) {1 \over \sqrt{4\pi g_-}}e^{i k_{+} x+i\varphi_-}+O(g_\pm^{-3/2})~~,
\end{equation}
where $\varphi_{\pm}=g_{\pm}+\pi/4$. 
Contributions $\phi_{\mbox{\tiny{tail}}}(x)$ are 'tails' whose amplitudes decay as 
$g_\pm^{-1/2}$.  As follows from (\ref{2.13}), $|g_\pm|\sim L/\lambda$ where  $\lambda$ is a wave length and $L$ is a distance 
related to position of the observer with respect to the string trajectory.

Physical effects in the far-field zone are interesting for the distant observers. Right observers crossing ${\cal H}_+$
interpret (\ref{2.17a}) as a refraction of the left wave on ${\cal H}_-$.
The surfaces $f_{\pm}(x)=0$ determine boundaries of diffraction of right and left parts of the wave behind the string.
The normal vectors $n_{\pm}$ to these surfaces  
$$
df_{\pm}=n_{\mu}dx^{\mu}~~,
$$
are orthogonal to the wave vectors
\begin{equation}
(n_\pm \cdot k_\pm)=0~~.
\end{equation}
The surfaces $f_{\pm}(x)=0$ intersect at the string world-sheet.

The domains of the diffraction  overlap. In the overlap region, $f_+>0$, $f_-<0$, $u>0$, the left 
and right waves interfere since the wave vectors $k_\pm$ are related by the nontrivial null rotation, $k_-=M(\omega)k_+$.
Thus, the null string leaves behind an interference wedge. This physical effect is similar to the effect of massive and null 
strings which leave behind the regions of overdensities of non-relativistic matter.

\begin{figure}
	\includegraphics[height=7cm]{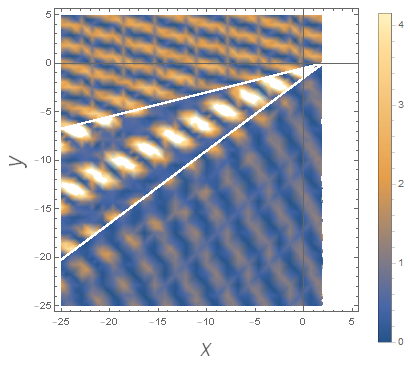}
		\includegraphics[height=7cm]{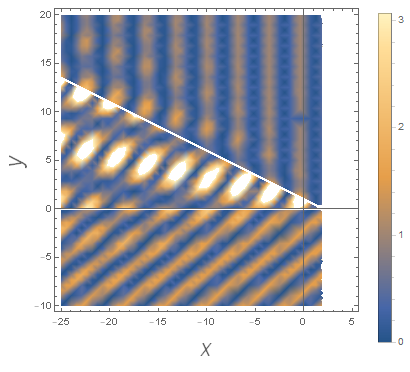}
	\caption{The energy density of a real scalar field  is shown for the string at the moment $t_0$.  The string is stretched along the $z$ axis, orthogonal to plane of the Figure, and is located at $x=t_0=2$, $y=0$;  $\omega=1/2$. For the left figure: $k_v>0$,  $k_{-y}=k_y-2\omega k_v>0$. For the right figure: $k_y=0$, $\cos \phi_{int}=(1+\omega^2)^{-1/2}$. }
	\label{fig1a}
\end{figure}

To demonstrate the existence of the overlap region we fix the moment $t=t_0$, put $k_+=k$, and suppose that $k_y>0$, 
$k_{-y}=k_y-2\omega k_v>0$. In coordinates $x$ and $y$ 
conditions $f_+>0$, $f_-<0$, $u>0$ look as
\begin{equation}\label{2.18}
x_-(y)<x<x_+(y)~~,~~x<t_0~~,
\end{equation}
\begin{equation}\label{2.19}
x_+(y)=t_0+{2k_v y\over k_y}~~,~~x_-(y)=t_0+{2k_v y \over k_y-2\omega k_v}~~.
\end{equation}
It is clear that conditions (\ref{2.18}) hold for $y<0$.  The angle of the  interference wedge, $\varphi_{\mbox{\tiny{intf}}}$,
 can be defined as the angle between
the lines $x=x_\pm(y)$
\begin{equation}\label{2.20}
\cos \varphi_{\mbox{\tiny{intf}}}={k_y(k_y-2\omega k_v)+4k_v^2 
\over (k_y^2+4k_v^2)^{1/2}((k_y-2\omega k_v)^2+4k_v^2)^{1/2}}~~.
\end{equation}
One can see that $\varphi_{\mbox{\tiny{intf}}}=O(\omega^2)$ at small $\omega$. The interference wedge exists at $k_y=0$ when 
$\cos \varphi_{\mbox{\tiny{intf}}}=(1+\omega^2)^{-1/2}$.

To illustrate this effect we evaluate the energy density $T_{00}(x)$, as measured by right observers for a real scalar field 
$\phi(x)=\Re\bigl(\phi_+(x) +\phi_-(x)\bigr)$, where $\phi_{\pm}$ are defined by  (\ref{2.15a}).
Fig. \ref{fig1a} shows $T_{00}(x)$ at the moment $t_0=2$ in the $(x,y)$ plane orthogonal to the string for the string parameter 
$\omega=0.5$. 	All coordinates are given in  dimensionless units (multiplied by $k_v=1$).  

We now return to solutions for EM waves. Solutions for incoming waves of the form 
$e^{-ik\cdot x}$ can be obtained from  (\ref{2.15a}), (\ref{2.15b}) by changing sign of the momentum.
As one can check by using  (\ref{2.15a}), (\ref{2.15b}) the solution is the complex conjugate of the solution for $e^{ik\cdot x}$.
With these remarks scattered EM  waves look (for $k_v>0$) as follows:
\begin{equation}\label{2.21a}
A^{\pm}_\mu(x)=\Re\left\{\left[\theta(\pm f_{\pm})+\varepsilon(\pm f_{\pm}) G(g_\pm)\right]~E^\pm_{\mu}\exp(i k_{\pm} x)\right\}~~,
\end{equation}
where the polarization vectors $E^\pm_{\mu}$ are defined by (\ref{2.23a}).
The interpretation of these results is the following. An observer inside the interference wedge sees the two waves with the same 
polarizations. The waves appear to come from distant sources which move with respect to each other.
This property is a combination of two effects known for moving cosmic strings, the lensing effect and the
creation of overdensities of matter.

\section{EM field of a point charge near null string}\label{ED}
\setcounter{equation}0

Consider a null string which moves near a point electric charge. Let $a$ be an impact parameter between the string and the charge.
Without loss of the generality
we suppose that the charge is at rest at a point with coordinates $x_e=z_e=0,y_e=a>0$. As we see, the string creates 
EM self-force which may change the velocity of the charge. We neglect this effect in the considered approximation and assume that
position of the charge remains fixed.
The corresponding current in (\ref{2.1})  is $\bar{j}^\mu(x)=e\delta^{(3)}(\vec{x}-\vec{x}_e) u^\mu$, with $u^\mu=\delta^\mu_0$.
The incoming field below $\cal H$ is
\begin{equation}\label{3.9}
\bar{A}_\mu(x)=-{e \over 4\pi}{\delta_\mu^0 \over \sqrt{x^2+(y-a)^2+z^2}}~~,
\end{equation} 
$x=(v-u)/2$.
Since the considered particle moves freely and crosses ${\cal H}_+$, in the $R$-chart the 4-velocity
of the particle is continuous. So do components of the current,
\begin{equation}\label{3.10}
j_\mu(x)=\bar{j}_\mu(x)~~,~~u>0~~.
\end{equation} 
Therefore the inhomogeneous part of the solution (\ref{3.7}) is taken as
\begin{equation}\label{3.11}
A_{I,\mu}(x)=\bar{A}_\mu(x)~~.
\end{equation} 
The homogeneous part is the solution to (\ref{3.8b}) with the following initial data:
\begin{equation}\label{3.12} 
a_{H,b}({\bf x})\mid_{{\cal H}_+}=0~~,~~a_{H,b}({\bf x})\mid_{{\cal H}_-}=
M_{b}^{~c}(\omega)\bar{a}_c(\bar{\bf x})-\bar{a}_b({\bf x})~~,
\end{equation} 
\begin{equation}\label{3.12b}
\bar{a}_b=- {e \over 8\pi}{\delta_b^v \over \sqrt{v^2/4+(y-a)^2+z^2}}~~.
\end{equation}

The homogeneous solution at $u>0$ can be represented as, see details in Sec.~\ref{App2},
\begin{equation}\label{3.18}
A_{H,\mu} (x)=-{e \over 8\pi^3} \int d\Omega~\Re\left[{b_\mu \over x^\nu ~m_\nu+ia\varepsilon }\right]
~~,
\end{equation} 
where integration goes over a unit sphere $S^2$, with coordinates $\Omega=(\theta,\varphi)$ and standard measure 
$d\Omega=\sin\theta d\theta d\varphi$.
The notations used in (\ref{3.18}) are the following:
\begin{equation}\label{3.19}
b_v=-\frac 12\cos\varphi \left(g^{-1}(\Omega,\omega)-g^{-1}(\Omega,0)\right)
~~,~~b_y=\omega \cos\varphi g^{-1}(\Omega,\omega)~~,~~b_z=0~~,
\end{equation} 
\begin{equation}\label{3.21}
g(\Omega,\omega)=e^{i\theta}+\omega \sin\theta \cos\varphi~~,~~\varepsilon=2\sin^2\theta\cos\varphi~~.
\end{equation} 
The vector $m_\mu$ is null, $m^2=0$,
\begin{equation}\label{3.22}
m_u=1-\sin^2\theta\cos^2\varphi~,~m_v=\sin^2\theta\cos^2\varphi~,~m_y=\sin 2\theta \cos\varphi~~,~~
m_z=\sin^2\theta\sin 2\varphi~.
\end{equation} 
The $A_u$ component is defined by the gauge condition $\partial A=0$, which is equivalent to  $b_\mu m^\mu=0$ and yields:
\begin{equation}\label{3.19u}
b_u={m_yb_y-2m_u b_v \over 2m_v}~~.
\end{equation} 

One can use (\ref{3.18}) to calculate electric, $\vec{E}^H$, and magnetic, $\vec{H}^H$,  fields created by the null string, 
\begin{equation}\label{3.23}
\vec{E}^H=\vec{\partial}A_{H0}-\partial_0\vec{A}_H~~,~~\vec{H}^H=[\vec{\partial}\times \vec{A}_H]~~. 
\end{equation} 
The total EM field above $\cal H$ is $\vec{H}=\vec{H}^H$, $\vec{E}=\vec{E}_C+\vec{E}^H$, where $\vec{E}_C$ is the Coulomb field of the charge
in the absence of the string.

As is known  \cite{Linet:1985}, a point charge, which is at rest near a massive straight string, 
experiences a self-force acting in the direction away from the string,
$F\sim G \mu e^2/r^2$, where $\mu$ is the tension of the string and $r$ is the distance between the charge and the string.
One of the analogous effects is a self-force of the charge in the presence of the null string. The self-force is determined by 
the electric field induced by the string, $\vec{F}(t,x_e)=e\vec{E}^H(t,x_e)$.  This force at short times is $F\sim \omega e^2/a^2
\sim GE e^2/a^2$, which is analogous to the self-force created by a massive string. At large times $t\gg a$ the self-force vanishes.

At large times numerical simulations for the potential $A_{H,0}$, 
and corresponding electric and magnetic fields  are shown on 
Fig.2 and  Fig.3, respectively. We take coordinates in (\ref{3.18}) as $x^0=t,x^i=rn^i$, where $n^i$ is a unit vector, and study 
potential and fields as functions  of arguments $r/t$ and $a/t$, at different fixed $t$. The component $E^H_y$ behaves similarly to $E^H_x$. Components $E_z^H$, $H_x^H$, $H_y^H$ are negligibly small.
As follows from these results, perturbation of EM field induced by the null string 
behaves as a EM pulse which moves away from the charge in different directions. 
The width of the pulse is 
determined by the impact parameter $a$. Such pulses can be specific experimental signatures of null strings moving near charged objects.

\begin{figure}[t]
	\label{VectP}
	\includegraphics[width=8cm]{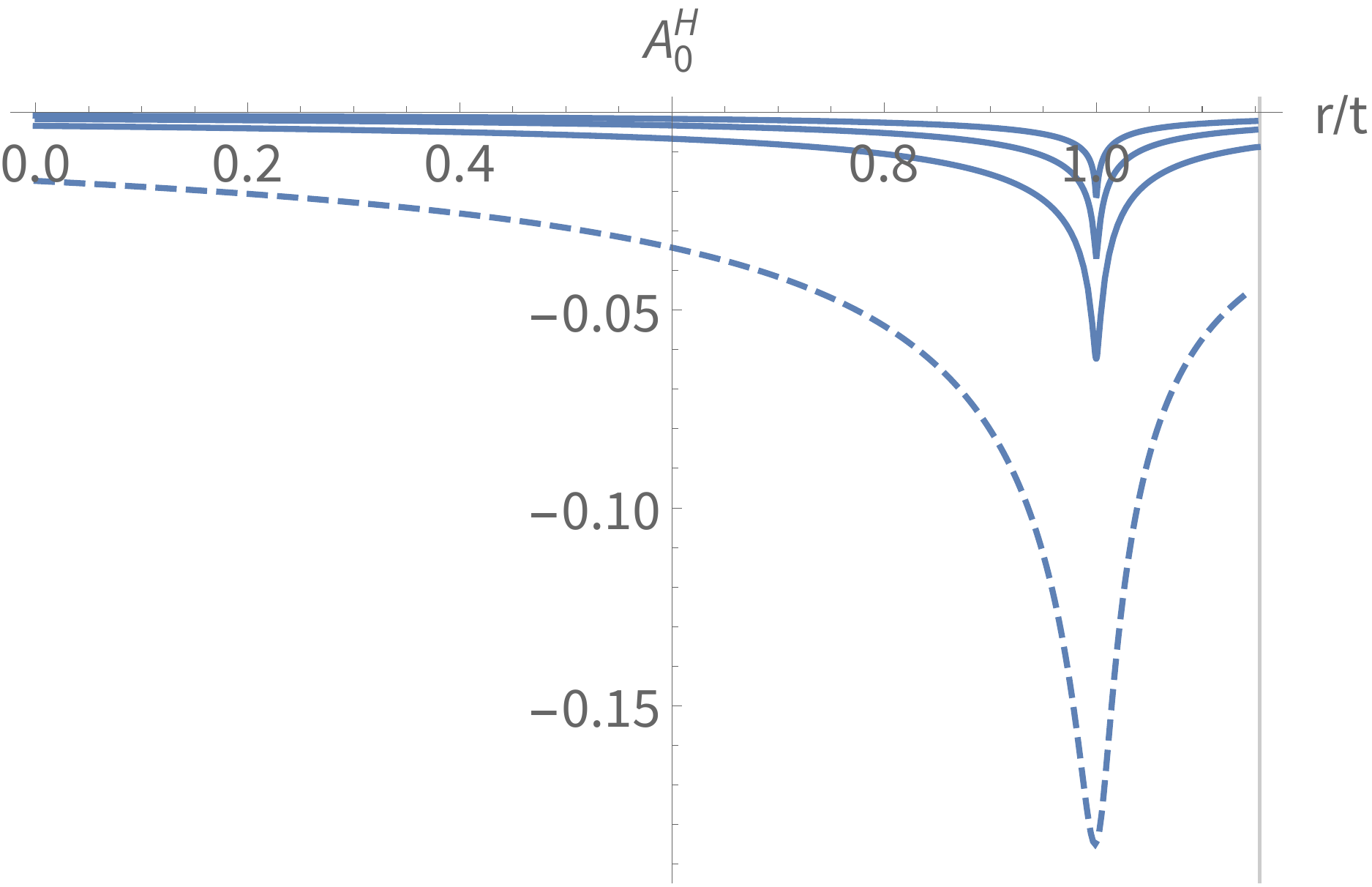} 
	\caption{The vector potential $A_{H,0}$   for $t/a=10, 50, 100, 200$ (from bottom to top) is plotted as a function of $r/t$. The largest amplitude of a pulse corresponds to $t/a=10$ (dashed curve). Here $\omega=1$, $a=1$, and $e=4\pi$. The observer 
coordinates are $(x=r\cos\pi/6, y=r\sin\pi/6, z=0)$. The string horizon is at  $r/t=1.154$.}
\end{figure}

\begin{figure}
	\label{EMf}
	\includegraphics[width=8cm]{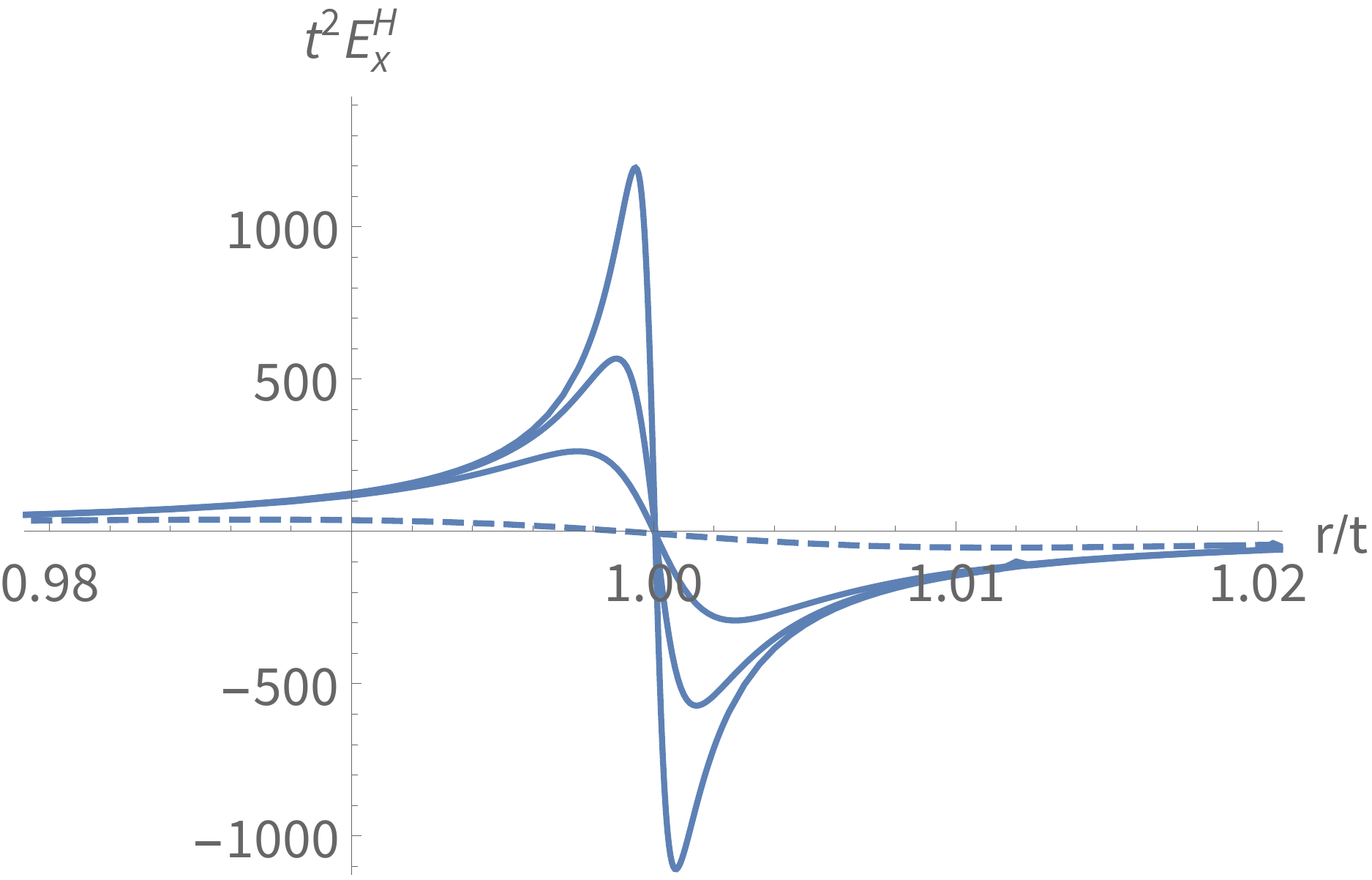} 
	\includegraphics[width=8cm]{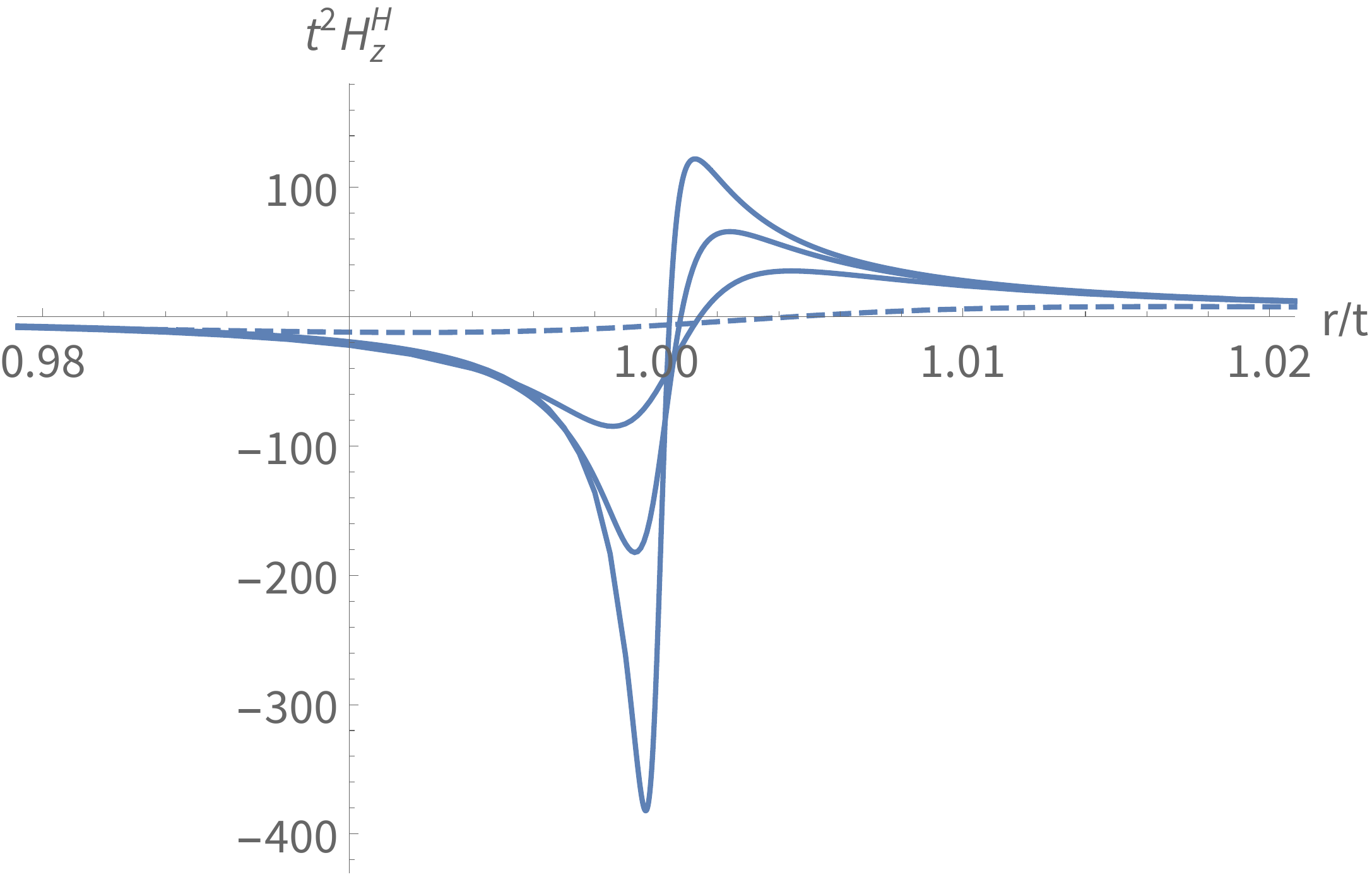} 
	\caption{The dimensionless functions $t^2 E^H_x$ and $t^2 H^H_z$ are plotted for $t/a=10$ (dashed), $50,\, 100, \,200$.   The observer's  coordinates are $(x=r\cos\pi/6, y=r\sin\pi/6, z=0)$,  $\omega=1$, $a=1$, and $e=4\pi$.}
\end{figure}

\begin{figure}
	\includegraphics[width=16cm]{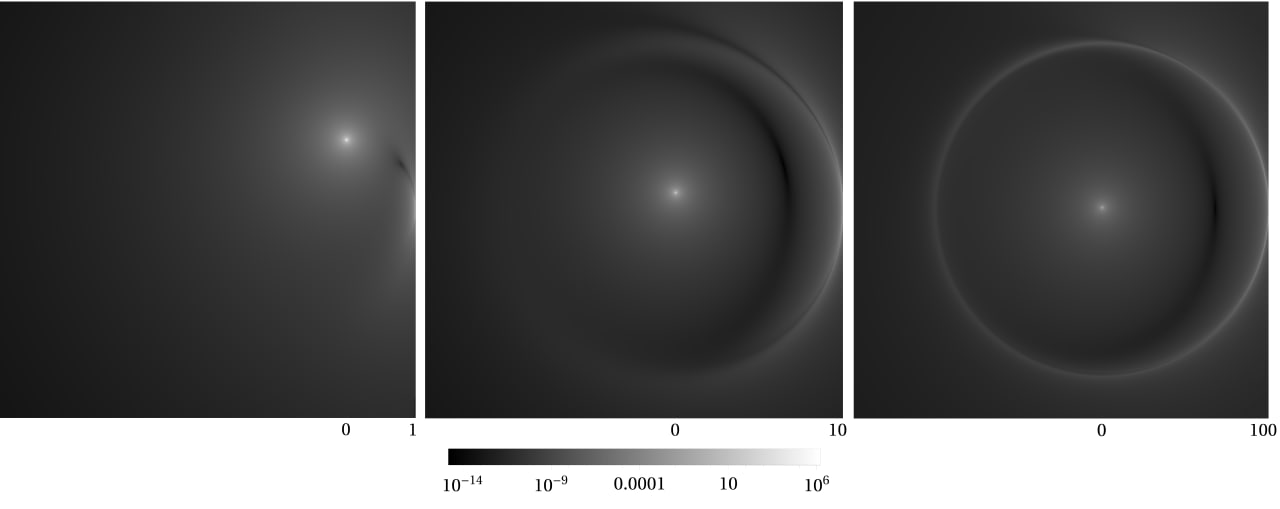} 
	\caption{The electromagnetic energy density of the system "charge + null string" is shown  in logarithmic scale at $t=1,\,10,\,100$. The energy density is computed in the plane $z=0$, the string moves from left to right, $\omega=1$, $a=1$, $e=1$. The string 
horizon is the right vertical sides of figures.}
	\label{Edens2}
\end{figure}

The energy of the EM field inside the sphere of the radius $R$ with the center at the point $x^i=0$ is
\begin{equation}\label{3.24}
E(R,t)=-\int_{r<R} d^3x T^0_0~~, 
\end{equation}
where $T^\mu_\nu$ is the stress-energy tensor of the EM field,
\begin{equation}\label{3.25}
T^\mu_\nu=-F^{\mu\alpha}F_{\nu\alpha}+\delta^\mu_\nu \frac 14 F^{\alpha\beta}F_{\alpha\beta}~~.
\end{equation}
The total energy density of EM field, $T_{00}$, as measured in the frame of reference where the charge is at rest, is shown on  Fig.\ref{Edens2}.

The conservation law implies that
\begin{equation}\label{3.26}
\partial_t E(R,t)=\int d\Omega~R^2 S(t,R,\Omega)~~, 
\end{equation}
\begin{equation}\label{3.27}
S(t,R,\Omega)=T_0^r(t,R,\Omega)=-F_{0i}F^{ri}~~. 
\end{equation}
Numerical results for the distribution of the energy flow $S(t,R,\Omega)$  at large times $t\sim r$ and $r \gg a$ 
are presented on  Fig.~5.
Simulation shows that the maximal pulse follows the string.

\begin{figure}
	\label{EFlow}
	\includegraphics[width=8cm]{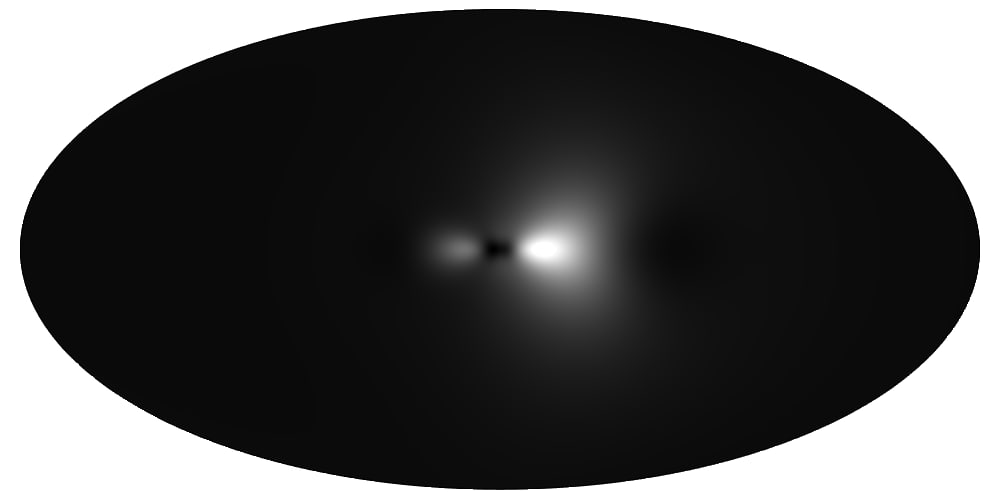} 
	\includegraphics[width=8cm]{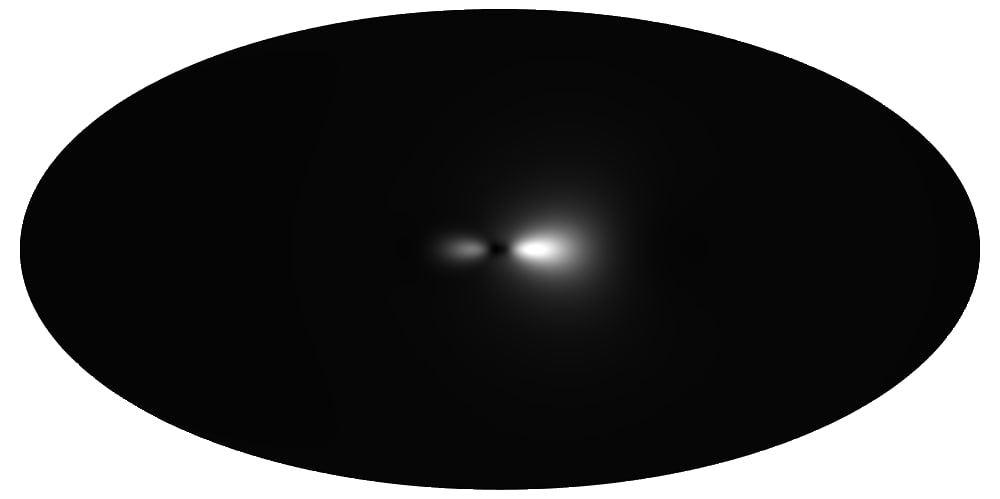} 
	\caption{The Molweide projection of the EM energy flow for $t/a=100$ and $R/a= 99.6$ (left), $99.8$ (right). Here $\omega=1$, $a=1$, $e=1$. The equator corresponds to $z=0$ plane. The string position is the central vertical line in the projection plane.}
\end{figure}

The maximum of the pulse is located on the light cone $r=t$, where $r$ is approximately the distance to the charge. The reason is that the denominator in the integral in (\ref{3.18}) has a minimum on the 
surface $x^\mu m_\mu=0$ which is a light cone with a future directed null normal vector $m$.

\section{Fields on gravitational shock wave backgrounds}\label{appl}
\setcounter{equation}0

The metric of space-time of a straight null string which is stretched along the $z$-axis and moves along the $x$-axis 
is known to be of the following form:
\begin{equation}\label{i.1}
ds^2=-dv du - \omega |y| \delta(u)du^2+dy^2+dz^2~~,
\end{equation} 
where $\omega$ is defined by (\ref{i.2}). This string space-time is locally flat \cite{Barrabes:2002hn}, except the world-sheet, where 
the $uu$ component of the Ricci tensor  has a delta-function singularity.
The delta-function in (\ref{i.1}) appears as a result of the null holonomy \cite{vandeMeent:2012gb}, \cite{Fursaev:2017aap}.

Geometry  (\ref{i.1}) is a particular example of gravitational shockwave backgrounds:
\begin{equation}\label{i.3}
ds^2=-dv du +f(y)\delta(u)du^2+\sum_i dy_i^2~~, ~~i=1,...n~~.
\end{equation}
Shockwaves (\ref{i.3}) are exact solutions of the Einstein equations sourced by a stress energy tensor localized at $u=0$ and 
having the only non-vanishing $uu$ component.
Another example of (\ref{i.3}) is the Aichelburg-Sexl solution corresponding to 
a gravitational field generated by a massless particle. 

As has been suggested by Penrose  \cite{Penrose:1972xrn}, to deal with (\ref{i.3})  one should cut $R^{1,n}$ along the hypersurface 
$u=0$ into two copies,
shift (supertranslate)  the $v$-coordinate of the upper copy ($u>0$) to $v-f(y)$ and 
glue the copies again. The shift of the $v$-coordinate can be also determined by working with a delta-function-like potential 
in wave-equations \cite{Klimcik:1988az}. This potential is generated by the $uu$-component of (\ref{i.3}).

In the case of null strings the Penrose prescription implies that $f(y)=-\omega |y|$.
By following Sec. \ref{charts}, consider a coordinate chart where the cut goes over the entire surface $u=0$ and choose the Penrose coordinate transformations:
\begin{equation}\label{1.4}
v=\bar{v}+\omega |y|~~,~~u=0~~.
\end{equation} 
According to (\ref{1.1}), on the left cut the coordinates are null rotated by the 'angle' $\omega/2$, while on the 
right with $-\omega/2$.  Since the relative null rotation on the right and left cuts is by angle $\omega$, 
condition  (\ref{1.4}) is equivalent to (\ref{1.3a}) or (\ref{1.3d}). 

Field theory near shockwaves is an interesting research subject. Previous results include calculations of the S-matrix for scattering scalar waves on shockwaves created by massless particles \cite{tHooft:1985NPB}, \cite{tHooft:1987PLB} and on generic gravitational shockwaves (\ref{i.3}), see \cite{Klimcik:1988az}, \cite{Lousto:1990wn}. 
In the last decade the interest to shockwaves has been related to black hole formation in high energy particle collisions.

Our approach can be used to describe solutions of hyperbolic PDE wave equations on gravitational shockwave backgrounds.
It is convenient to write the Penrose transition condition for coordinates as
\begin{equation}\label{5.1}
x^\mu=(\bar{x}^\mu-\zeta^\mu(\bar{x}))\mid_{u=0}~~,~~\zeta^\mu=\delta^\mu_v f(y)
\end{equation} 
Suppose $|f(y)|\ll 1$. Then the transition condition for a field $\phi$, which generalizes (\ref{1.3cs}), is
\begin{equation}\label{5.2}
\hat{\phi}(x)=(\phi(x)+{\cal L}_\zeta \phi(x))\mid_{u=0}~~,
\end{equation} 
where $\phi(x)$ is the value of the field in the absence of the shockwave, and ${\cal L}_\zeta \phi$ is the Lie derivative of the 
field generated by the vector field in (\ref{5.1}).  Solutions to field equations are determined by a Cauchy problem with conditions
(\ref{5.2}) at $u=0$ and can be constructed in the same way as for EM fields in the presence of a null string.

\section{Summary and perspectives}\label{sum}
\setcounter{equation}0

In this work we suggested a method to describe free classical fields on gravitational background of a straight null string 
and, more generally, on shockwave backgrounds.  Applications of the method have been focused on scalar and electromagnetic fields.

We described two new physical effects: scattering of plane electromagnetic waves by null strings and generation of EM fields  by null strings passing 
by near point charges. Both effects can be used in astrophysical observations to search for null cosmic strings.

There are several avenues where the method can be applied. By proceeding with EM phenomena it is interesting to study effects induced by null cosmic strings near  objects which possess strong EM fields, such as black holes and neutron stars.

One can extend the results of Sec. \ref{ED} to linearized Einstein equations and  study gravitational perturbations generated by null 
cosmic strings moving near massive bodies. Like in case of EM fields we expect pulses of gravitational radiation and gravitational self-force effects
generated by the strings.

It should be pointed out that the suggested method allows one to introduce different Green's functions on null string space-times,
and, therefore, to pave a way to quantum field effects related to null strings.

We are planning to return to some of this topics in forthcoming works.

\section{Acknowledgments}

The authors are grateful to V. Tainov for the help with some technicalities and numerical simulations.
This research is supported by Russian Science Foundation grant No. 22-22-00684, 
https://rscf.ru/project/22-22-00684/.

\bigskip
\bigskip
\bigskip

\newpage
\appendix

\section{The $D$-function, solution for scattered scalar wave}\label{App1}
\setcounter{equation}0

We derive the $D$-function for the massless field, $m=0$ in (\ref{1.7}),
by using the following representation: 
\begin{equation}\label{a1}
D(u,{\bf x})={1 \over (2\pi)^3}\int  d^3 p \, e^{ip\cdot x}~~,
\end{equation}
where $p\cdot x=p_{\mu}x^{\mu}=p_u u+p_v v +p_y y +p_z z$, $d^3p=dp_vdp_ydp_z$
\begin{equation}\label{a2}
p_u=\frac{p_y^2+p_z^2}{4p_v}~~.
\end{equation}
It is convenient to slightly shift the integration contour over $p_v$ in (\ref{a1}) to the lower part of the complex $p_v$-plane to ensure convergence of the integrals over $p_y$ and $p_z$. A simple formula  
\begin{equation}\label{1.7a}
D(x)={1 \over \pi}{\partial \over \partial v}\delta (x^2)~~,
\end{equation}
where $x^2=x^\mu x_\mu=-uv+y^2+z^2$, can be easily derived when one first integrates over $p_y$, $p_z$ and then over $p_v$.

It is instructive to check that (\ref{1.7a}) satisfies  (\ref{1.7}). A direct calculation yields
\begin{equation}\label{a8}
\Box D(x)={4 \over \pi u}(3\delta''(x^2)+x^2\delta'''(x^2))~~.
\end{equation}
If $f(r)$ is a test function on the line,
\begin{equation}\label{a9}
\int dr f(r) (3\delta''(r)+r\delta'''(r))=-\lim_{r\to 0} (rf'''(r))~~.
\end{equation}
Therefore the right hand side of (\ref{a8}) vanishes for functions which are analytical at $x^2=0$. 
Define now the functional
\begin{equation}\label{a10}
F[u,\chi]=\int d^3x D(u,{\bf x})\chi({\bf x})~~
\end{equation}
acting, say, on a $L^2$ space of test funtions $\chi({\bf x})$.  We need to prove that $F[0,\chi]=\chi(0)$. By 
virtue of (\ref{1.7a})
\begin{equation}\label{a11}
F[u,\chi]=-{1 \over \pi}\int d^3x \delta(x^2)\partial_v\chi({\bf x})~~.
\end{equation}
If $\rho,\varphi$ are polar coordinates in $z,y$ plane
\begin{equation}\label{a12}
F[u,\chi]=-{1 \over \pi}\int dv \rho d \rho  d\varphi~ \delta(\rho^2-uv)\partial_v\chi(v,\rho,\varphi)=
-{1 \over 2\pi}\int_0^\infty dv \int_0^{2\pi}d\varphi~\partial_v\chi(v,\sqrt{uv},\varphi)~~.
\end{equation}
Since ${d \over dv}\chi(v,\sqrt{uv},\varphi)=\partial_v\chi(v,\sqrt{uv},\varphi)+\frac{1}{2}\sqrt{\frac{u}{v}}\partial_\rho \chi(v,\sqrt{uv},\varphi)$
\begin{equation}\label{a13}
F[u,\chi]=\chi(0)+{\sqrt{u} \over 4\pi}\int_0^\infty \frac{dv}{\sqrt{v}} \int_0^{2\pi}d\varphi~\partial_\rho\chi(v,\sqrt{uv},\varphi)~~.
\end{equation}
The last term in the right hand side of (\ref{a13}) vanishes as $u\to 0$, if $\chi$ is analytical at $\rho=0$.

We now present derivation of basic formulas (\ref{2.15a}), (\ref{2.15b}) for scattering by the null string of a scalar
wave. 
The $\phi_\pm$ parts of the scattered wave (\ref{2.12}) are defined by (\ref{1.6}). One gets with the help of (\ref{a1}) 
\begin{equation}\label{a3}
\phi_{\pm}(u, {\bf x})=\frac{1}{(2\pi)^3}\int d^3p\, e^{i {\bf p}{\bf x} }
\exp \left({i \frac{p_y^2+p_z^2}{4 p_v}u}\right)\int d^3 x'\, e^{-i {\bf p} {\bf x'}}\,\hat{\phi}_{\pm}({\bf x'})\theta(\pm y')~~,
\end{equation}
where ${\bf p} {\bf x'}=p_vx^v+p_yy+p_zz$ and 
\begin{equation}\label{a4}
\hat{\phi}_{\pm}({\bf x})=e^{ik_\pm\cdot x}\mid_{{\cal H}_\pm}~~,~~k^\mu_+=\bar{k}^\mu~~,~~k_-^\mu=M^\mu_{~\nu}\left(\omega\right)\bar{k}^\nu~~.
\end{equation}
To simplify notations we put $\bar{k}^\mu=k^\mu$ in what follows.
One can perform integration in (\ref{a4}) first over ${\bf x'}$, then over $p_z$ and $p_v$ and get
\begin{equation}\label{a5}
\phi_{\pm}(u,{\bf x})=\pm I_\pm(k,x) \exp i\left(k_v v+k_z z+\frac{k_z^2}{4k_v}u-\frac{k_v}{u}y^2\right)~~,
\end{equation}
\begin{equation}\label{a6}
I_\pm(k,x)={1 \over 2\pi i}
\int\limits_{-\infty}^{\infty} 
\frac{ d p \,e^{i\frac{p^2}{4uk_v}}}{p- (f_\pm \pm i \epsilon)}~~,
\end{equation}
where $p=up_y$, and $f_\pm =f(k_\pm,x)$ are defined in (\ref{2.13}).  Factors $I_\pm$ appear as a result of integration over $y$.

It is convenient to rewrite  (\ref{a6}) by changing contour in the complex $p$-plane. Since $u>0$,  the integration contour over $p$ can be rotated  by the angle $\pi/4$  either counter-clockwise, if $k_v>0$, or clockwise, if $k_v<0$. This yields
\begin{eqnarray}
\label{a7}
&&I_\pm(k,x)=\pm\left[\theta(k_v)\theta(\pm f_\pm)+\theta(-k_v)\theta(\mp f_\pm)\right]e^{i\frac{f_{\pm}}{4  k_v u}}  \\[6pt]
&&+{\theta(k_v) \over 2\pi i}\int_{-\infty}^{\infty}{dt \over t-e^{-i\pi/4}f_\pm}\exp\left(-{t^2 \over 4 k_v u}\right)+
{\theta(-k_v) \over 2\pi i}\int_{-\infty}^{\infty}{dt \over t-e^{i\pi/4}f_\pm}\exp\left({t^2 \over 4k_v u}\right)~~.
\nonumber
\end{eqnarray}
The first two terms in (\ref{a7}) appear when the rotation of the contour meets poles. Eqs. (\ref{2.15a})-(\ref{2.16}) 
follow from (\ref{a7}) if one replaces $t$ in the integrals with $f_\pm t$ and takes into account that $k_u=(k_y^2+k_z^2)/{4k_v}$.

\section{Homogeneous solution for point electric charge}\label{App2}
\setcounter{equation}0

Here we present computations for Sec. \ref{ED}. Define
\begin{equation}\label{3.14}
f(v,y,z)=\theta(-y) {e \over 4\pi}{1 \over \sqrt{v^2/4+(y-a)^2+z^2}}~~,
\end{equation} 
By using (\ref{1.1a}), (\ref{3.12b}) the Cauchy data (\ref{3.12}) can be written as
\begin{equation}\label{3.15u}
a_{H,v}({\bf x})=-\frac 12(f(v-2\omega y, y, z)-f(v,y,z))~~,
\end{equation}  
\begin{equation}\label{3.15y}
a_{H,y}({\bf x})=\omega f(v-2\omega y, y, z)~~,~~a_{H,z}({\bf x})=0~~.
\end{equation} 
Let $\Phi_\omega(u, {\bf x})$ be a solution at $u>0$ of the 
following problem:
\begin{equation}\label{a2.1}
\Box \Phi_\omega(u, {\bf x})=0~~,~~\Phi_\omega(0, {\bf x})=f(v-2\omega y, y, z)~~.
\end{equation} 
Then the homogeneous part of non-zero components of the vector-potential are
\begin{equation}\label{a2.2}
A_{H,v}(x)=-\frac 12(\Phi_\omega(x)-\Phi_0(x))~~,
\end{equation} 
\begin{equation}\label{a2.4}
A_{H,y}(x)=\omega \Phi_\omega(x)~~.
\end{equation} 
The solution to (\ref{a2.1}) can be written as
\begin{equation}\label{a2.5}
\Phi_\omega(x)={1 \over (2\pi)^{3/2}}\int d^3k~ e^{ik \cdot x}~\tilde{f}_\omega({\bf k})~~,
\end{equation} 
\begin{equation}\label{a2.6}
\tilde{f}_\omega({\bf k})={1 \over (2\pi)^{3/2}}\int d^3x e^{-i{\bf k}{\bf x}}f(v-2\omega y, y, z)~~,
\end{equation} 
where $d^3x=dvdydz$, and we use the same notations as in Sec.\ref{App1}.  A straightforward computation yields
\begin{equation}\label{a2.7}
\tilde{f}_\omega({\bf k})={e \over (2\pi)^3}
{e^{-a\sqrt{(2k_v)^2+k_z^2}} \over \sqrt{(2k_v)^2+k_z^2} 
\left( \sqrt{(2k_v)^2+k_z^2}-i(k_y+2\omega k_v)\right)}~~.
\end{equation} 
It is convenient to introduce in (\ref{a2.5}) spherical coordinates in the momentum space, $2k_v=k\sin\theta\cos\varphi$,
$k_z=k\sin\theta\sin\varphi$, $k_y=k\cos\theta$. By taking into account (\ref{a2.7}) the integration over $k$ in (\ref{a2.5})  can be performed,
\begin{equation}\label{a2.8}
\Phi_\omega(x)=-{e \over 8\pi^3} \int_0^{2\pi} d\varphi  \int_0^{\pi} \sin\theta d\theta~{1 \over g(\Omega,\omega)}~
{1 \over x^\mu\,m_\mu +ia\varepsilon}
~~,
\end{equation}
see notations (\ref{3.21}), (\ref{3.22}).  Equations (\ref{a2.2})-(\ref{a2.4}), (\ref{a2.8}) imply (\ref{3.18}),(\ref{3.19}).

\newpage
\bibliographystyle{unsrt}
%\bibliography{nullStr.bib}
%\end{document}

\end{document}